\documentclass[aps,pre,reprint,superscriptaddress,amsmath,amssymb]{revtex4-2}
\usepackage[utf8]{inputenc}
\usepackage[T1]{fontenc}
\usepackage{graphicx}
\usepackage{booktabs}
\usepackage{multirow}
\usepackage{hyperref}

\newcommand{\avg}[1]{\langle #1\rangle}

\begin{document}

\title{Block-Fitness Modeling of the Global Air Mobility Network}

\author{Giulia Fischetti}
\thanks{These authors contributed equally to this work.}
\affiliation{DMSN, Ca' Foscari University of Venice, 30172 Mestre, Italy}
\affiliation{IAMP, ZHAW School of Engineering, 8400 Winterthur, Switzerland}

\author{Anna Mancini}
\thanks{These authors contributed equally to this work.}
\affiliation{Physics Dept. and INFN, University of Rome Tor Vergata, 00133 Rome, Italy}
\affiliation{Enrico Fermi Research Center, 00184 Rome, Italy}

\author{Giulio Cimini}
\email{giulio.cimini@roma2.infn.it}
\affiliation{Physics Dept. and INFN, University of Rome Tor Vergata, 00133 Rome, Italy}
\affiliation{Enrico Fermi Research Center, 00184 Rome, Italy}
\affiliation{CNR-Institute for Complex Systems (ISC), 00185 Rome, Italy}

\author{Jessica T. Davis}
\affiliation{MOBS Lab, Northeastern University, Boston MA 02115, USA}

\author{Abby O. K. Leung}
\affiliation{MOBS Lab, Northeastern University, Boston MA 02115, USA}

\author{Alessandro Vespignani}
\affiliation{MOBS Lab, Northeastern University, Boston MA 02115, USA}

\author{Guido Caldarelli}
\email{guido.caldarelli@cnr.it}
\affiliation{CNR-Institute for Complex Systems (ISC), 00185 Rome, Italy}
\affiliation{DMSN, Ca' Foscari University of Venice, 30172 Mestre, Italy}
\affiliation{ECLT Ca' Bottacin, 30123 Venice, Italy}
\affiliation{LIMS, Royal Institution, London W1S4BS, United Kingdom}

\begin{abstract}
Accurate representations of origin-destination (OD) passenger flows on the World Air Transportation Network (WAN) are fundamental inputs to models of global mobility, epidemic risk, and infrastructure planning.
However, OD data are largely commercial and proprietary, therefore often inaccessible to the research community.
Here we introduce a generative model that treats air mobility as a stochastic process within a sparse maximum-entropy framework, informed by aggregate traffic volumes of individual locations and geographic regions.
The model accurately reconstructs both network structure and mobility flows, yielding epidemic spreading trajectories that closely match those obtained using real mobility data.
Furthermore, the method can be effectively implemented using open data on flight routes and airport traffic, reducing dependence on proprietary datasets.
Our approach provides a scalable, interpretable, and computationally efficient framework for scenario analysis and policy design in global mobility systems.
\end{abstract}

\maketitle

\section{Introduction}

Statistical physics provides a powerful framework to describe collective behavior in complex systems.
In recent years, this approach has extended beyond traditional domains to capture socio-economic phenomena, where large-scale patterns emerge from the interactions of many heterogeneous agents \cite{durlaf,demartino2006,castellano2009statistical,PERC2017,caldarelli2018physics,barthelemy2019,cimini2019statistical,bouchaud2025}.
A central example is epidemic spreading, in which underlying mobility networks shape the pathways of disease transmission across populations
\cite{balcan2009multiscale,brockmann2013hidden,tizzoni2014use,pastor2015epidemic,oliver2020mobile,sills2020,wesolowski2012,kostandova2024,LU20261}.
In particular, the World Air Transportation Network (WAN) drives global contagion dynamics by rapidly connecting distant regions
\cite{flahault1992method,grais2003assessing,hufnagel2004forecast,colizza2006role,hollingsworth2006}.

Despite their key importance in computational epidemiology \cite{lessler2009,wardle2023,NAP26599}, relevant datasets quantifying air passenger mobility are difficult to access. Openly-available data consist of scheduled flights and seat capacities \cite{barrat2004architecture,zanin2013modelling}, which describe the physical transportation infrastructure, but provide information only on direct flight connections and misestimate realized passenger flows \cite{johansson2011,eurosurv2019}. Indeed  traffic measured on individual legs tends to overestimate mobility importance at transit hubs and underestimate it at actual destinations.
Origin-destination (OD) datasets derived from air ticket sales instead capture how many individuals travel from a source population to a final destination, regardless of intermediate flight segments. Therefore they provide a more accurate representation of the population mixing underlying epidemic spread, improving large-scale forecasting performance \cite{nicolaides2012,gomez2019}. As such, their use has become standard in recent literature \cite{wang2018,davis2021,klasmer2024,wardle2024}.


However, OD datasets are typically proprietary, expensive for research purposes and difficult to share openly.
This limitation motivates the development of methods capable of reconstructing global air mobility from more accessible information.
Existing approaches infer OD flows using gravity-like models \cite{zipf1946hypothesis,grosche2007,barbosa2018human,simini2021deep,cabanastirapu2023human}, which combine demographic, socio-economic and geographic variables to estimate movement between locations.
Machine-learning approaches (see \cite{ehsani2024predicting} and references therein) extend this strategy by exploiting richer information sets, improving predictive performance at the expense of interpretability. While successful in many settings, these approaches rely on strong modeling assumptions and on the selection of exogenous informative covariates. Moreover, unless additional sparsity assumptions are introduced (such as conditioning on the topology extracted from flight schedules \cite{huang2013open,mao2015worldwide}), they typically produce dense or effectively fully connected systems---at odds with the observed sparsity of air connectivity arising from operational, economic, and geographical constraints.

Here we aim to overcome these limitations, by proposing a generative framework to reproduce both structure and flows of the OD global air mobility network using a minimal set of endogenous information.
To this end, we draw on network science approaches for reconstructing weighted networks.

Focusing on topology extracted from flight routes, previous studies show that the global air mobility network exhibits heavy-tailed degree distributions, small-world properties and a modular organization characterized by dense regional communities connected by long-range intercontinental routes \cite{guimera2005worldwide,diop2021revealing}. Several generative approaches have been proposed to reproduce these features. Preferential attachment modulated by geographic distance captures the emergence of hubs and local clustering \cite{guimera2004modeling}. Entropy-based spatial formulations assign connection probabilities as functions of node degree and distance, reproducing similar large-scale connectivity patterns \cite{bianconi2008entropy,bianconi2021information,picciolo2012role}.
However, these approaches largely neglect link weights, which are not trivially determined by connectivity \cite{barrat2004architecture}. Weighted maximum-entropy approaches that constrain passenger volumes at each location have not been explored, as they generate dense connectivity patterns analogous to gravity models \cite{cimini2019statistical}.

We instead build on a two-step reconstruction framework developed for economic networks \cite{squartini2018reconstruction}.
First, we infer the existence of connections through a maximum-entropy fitness model \cite{caldarelli2002scale,garlaschelli2004fitness}. Second, we assign passenger volumes conditional on those links through a density-corrected gravity mechanism \cite{cimini2015systemic}.
We extend this approach by introducing constraints that reflect the modular organization of global air mobility into geographic communities \cite{karrer2011stochastic,fronczak2013exponential}.
Specifically we use traffic-by-region at individual locations as fitness governing their connectivity (see Methods).
The resulting block-fitness model provides a probabilistic framework to generate synthetic realizations of the network while preserving analytical tractability.

We validate the proposed framework against empirical OD data and compare its performance with alternative formulations.
The model accurately recovers individual connections, traffic statistics and large-scale mobility patterns of the empirical network. When employed in metapopulation epidemic simulations, synthetic networks reproduce spreading trajectories that closely match those obtained on real data, including their spatial heterogeneity and temporal dynamics.
Importantly, the model is informed solely by aggregate mobility patterns, which we extract from a proprietary OD dataset, but can also be efficiently calibrated using open information on flight routes and airport passenger counts.
Beyond reconstruction performance, the framework remains interpretable and scalable while providing insight into the mechanisms underlying the formation of global air mobility networks, offering a practical tool for researchers and practitioners in network epidemiology.

\section{Results}

\subsection{WAN representation and properties}\label{airport_database}
\begin{figure*}[t]
    \centering
    \includegraphics[width=\textwidth]{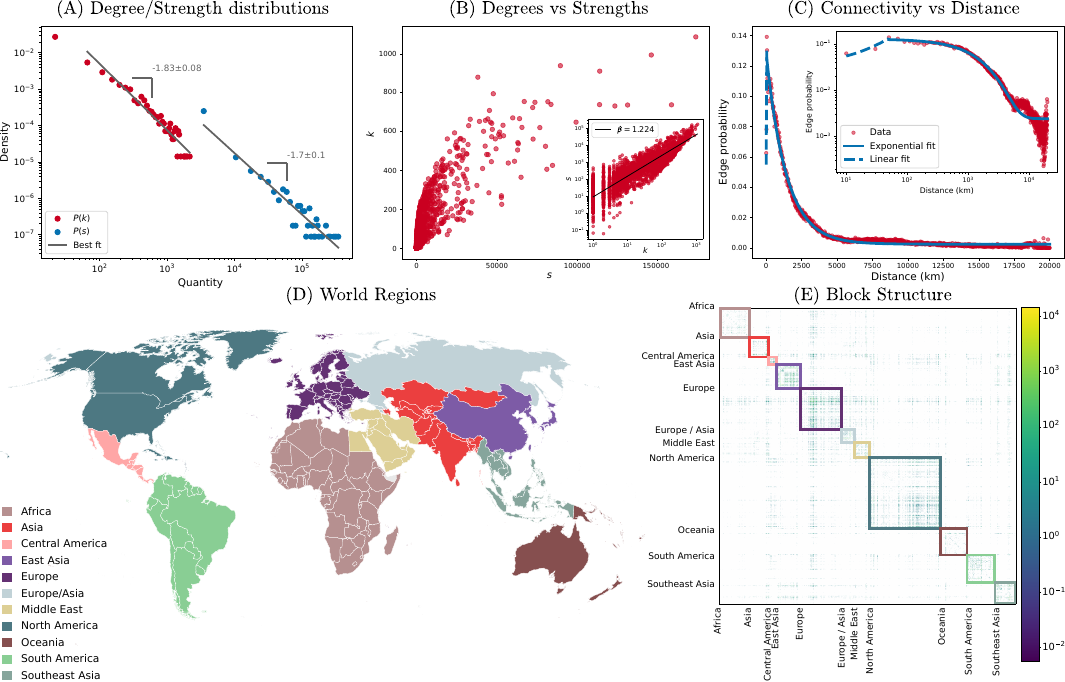}
    \caption{\textbf{Topological features of the WAN.}
    (A) Degree distribution with power-law fit $P(k)\sim k^{-\gamma_k}$ and exponent $\gamma_k=1.8(3)$, and strength distribution with power-law fit $P(s)\sim s^{-\gamma_s}$ and exponent $\gamma_s=1.(7)$ (50 data bins are used to represent each distribution).
    (B) Degree-strength relation of individual basins, with OLS fit $s\sim k^\beta$ and exponent $\beta=1.2(2)$. (C) Link probability as a function of geographic distance (1000 data bins of approximately 20~km width are used).
    The curve is fitted with a linear function $l(d) = m\cdot d+q$ up to a characteristic distance $d^*=50$~km and then with an exponential $l(d) = \exp[-a(d-d_0)]+b$.
    Best-fit parameters are $m=0.003$, $q=-0.0013$ for the linear term and $d_0=-2.(9)\cdot10^3$, $a=7.(1)\cdot10^{-4}$, $b=-2.(4)\cdot10^{-3}$ for the exponential decay ($R^2=0.99$).
    (D) World partition into regions: Africa, Asia, Central America, East Asia, Europe, Europe/Asia, Middle East, North America, Oceania, South America, Southeast Asia. (E) Community structure in the WAN adjacency matrix induced by the world partition, with modularity $M=0.67$.}
    \label{fig:basin_feat}
\end{figure*}

In this work, we define the WAN as the worldwide origin--destination (OD) air passenger network. We thus consider air-travel data from the Official Aviation Guide (OAG)~\cite{OAG}, which reports monthly counts of passengers traveling from an origin airport to their final destination, independently of intermediate layovers.
We group airports into ``basins'' by dividing the world into over 3200 geographic subpopulations generated using a Voronoi tessellation of the Earth's surface, where subpopulations are centered around major transportation hubs (details in Supplementary Methods). Passenger flows are then summed across each basin pair. We take the monthly basin networks from September to November 2019 and compute edge weights by averaging over the number of daily travelers. The resulting network is undirected, with link weight $w_{ij}$ given by the flux of passengers in both directions between basins $i$ and $j$.
For each basin $i$, the passenger volume is measured by the strength $s_i = \sum_j w_{ij}$ and the number of connections by the degree $k_i = \sum_j a_{ij}$ ($a_{ij}=1$ if $w_{ij}>0$ meaning that $i$ and $j$ are connected, and zero otherwise).
Degree and strength distributions are well approximated by power-laws with exponential cut-offs \cite{serafino2021true}, reflecting capacity constraints that limit the number of connections and passengers a single basin can accommodate (Fig.~\ref{fig:basin_feat}A). The relation between $k$ and $s$ is sub-linear, with some dispersion (Fig.~\ref{fig:basin_feat}B).
The distance $d_{ij}$ between basins $i$ and $j$ is computed using the Haversine formula applied to the centroids of the two basins.
Link occurrence does depend on geographical distance, as shown by the frequency of connections at different distance intervals (Fig.~\ref{fig:basin_feat}C).
Basins are partitioned into 11 geographical regions derived from the United Nations geoscheme and the World Bank regional classification (Fig.~\ref{fig:basin_feat}D).
Such partition induces a block structure in the network with very high modularity \cite{modularity} (Fig.~\ref{fig:basin_feat}E).

\subsection{Network reconstruction}
\label{link_reconstruction}

The central element of our modeling framework is the fitness ansatz, for which node strengths serve as parameters that control expected degrees. This assumption is justified when degrees are strongly correlated with strengths, as is the case for the WAN (Fig.~\ref{fig:basin_feat}B). Consistent with this premise, the model accurately reconstructs node degrees as ensemble expectations (Fig.~\ref{fig:rec_basin}A), while node strengths are reproduced by construction (see Supplementary Methods).
Concerning the inference of individual links, we measure the empirical link probability as a function of the model link probability $p_{ij}$, by binning $p_{ij}$ values and counting the fraction of linked $(i,j)$ pairs in each bin.
We obtain a very good agreement between the two quantities (Fig.~\ref{fig:rec_basin}B). Besides, the model recovers the weights of existing WAN connections as aggregate ensemble averages $\langle w_{ij}\rangle$ (Fig.~\ref{fig:rec_basin}C).

We further employ the model link probability $p_{ij}$ for a link prediction exercise that determines the outcome of a binary classification test on the existence of a link, given by $a_{ij}$.
Values of precision ($\sum_{ij}a_{ij}p_{ij}/\sum_{ij}p_{ij}$) and recall ($\sum_{ij}a_{ij}p_{ij}/\sum_{ij}a_{ij}$) of the classifier are equivalent in our case, since by construction model networks have on average the same number of links as the empirical one (see Methods).
We obtain a very high recall of 75\%, meaning that the model is able to correctly identify about three out of four existing links, despite the dataset being highly unbalanced with only a few ``positive'' existing links: the network density is 1.8\%, corresponding to the expected precision of a random classifier.

The model can also reconstruct the signature properties of the WAN. It captures the trend and the full dispersion of the $k$ vs $s$ relation (Fig.~\ref{fig:rec_basin}D).
It reproduces, to a large extent, the empirical dependence of link occurrence on geographic distance (Fig.~\ref{fig:rec_basin}E), despite not using any explicit spatial information. By construction, it also recovers the network's block structure (Fig.~\ref{fig:rec_basin}F).

\subsection{Spreading simulations}
\label{compartmental_model_simulations}

\begin{figure*}[t]
    \centering
    \includegraphics[width=\textwidth]{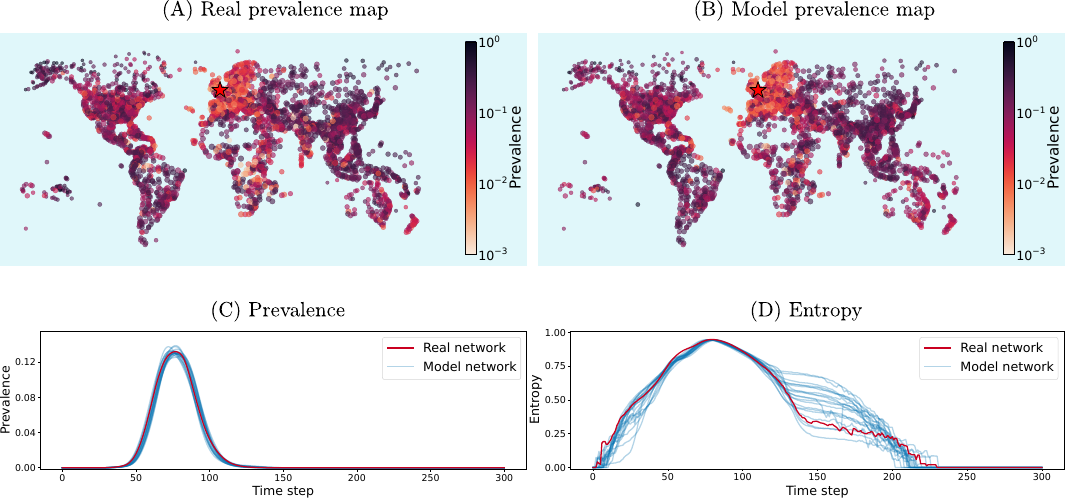}
    \caption{\textbf{Epidemic simulations starting from London.}
    (A-B) Prevalence map at simulation step $t=80$ for the empirical (A) and model (B) WAN.
    Each point represents a basin placed at its geographical coordinates, with marker size proportional to degree and color representing the prevalence (only basins with a prevalence of at least $10^{-3}$ are displayed).
    (C-D) Time evolution of the prevalence (total fraction of infected individuals) (C) and of the entropy of the prevalence distribution (D), for one simulation on the empirical network and 20 simulations on model networks.}
    \label{fig:meta_results}
\end{figure*}

Next we evaluate whether the block-fitness model reproduces epidemic dynamics on the empirical WAN.
We use a stochastic metapopulation model for global epidemic spreading, where the compartmental equations in each basin are coupled through the passenger flows set by the WAN travel matrix \cite{balcan2009multiscale,Desiderio_2022}
(see Methods for further details).
Figure~\ref{fig:meta_results} illustrates the spatiotemporal dynamics of contagion on both the real and model WAN, for a representative case in which the epidemic outbreak originates in London.
Prevalence maps (Fig.~\ref{fig:meta_results}A and Fig.~\ref{fig:meta_results}B for the real and model networks, respectively) show the global distribution of basin-level prevalence $i_m(t)$ at a fixed time, where $i_m(t)$ denotes the fraction of infected individuals in basin $m$ at time $t$ relative to its total population.
The temporal evolution of the total prevalence $i(t)$---defined as the fraction of infected individuals worldwide (Fig.~\ref{fig:meta_results}C)---exhibits the characteristic epidemic peak.
To quantify the spatial heterogeneity in disease prevalence, we compute the normalized spatial distribution $\rho_m(t)=i_m(t)/\sum_n i_n(t)$ and its associated entropy $H(t)\propto -\sum_m \rho_m(t)\log \rho_m(t)$. This quantity vanishes when the epidemic is fully localized within a single basin and is maximized when the epidemic is homogeneously spread across the globe \cite{colizza2006role} (a condition that typically occurs in correspondence of the epidemic peak, Fig.~\ref{fig:meta_results}D).

Although visual inspection already suggests strong agreement between contagion dynamics on real and model networks (the prevalence maps and the epidemic peak are very well reproduced), we consider several complementary metrics to quantitatively compare simulations on real and model WANs. We measure the overall mismatch between the prevalence curves using the area between curves and their L2 distance, while we capture differences in outbreak timing, intensity and duration by measuring the difference in peak time, peak magnitude and epidemic length. Further we compute the Root Mean Square Error (RMSE) of the basin-level prevalence values $i_m(t)$, averaged over time, and the L2 distance between the entropy curves.
Table~\ref{tab:epidemic_model_B_vs_real_variability} reports values of these quantities, averaged over several simulations on empirical or model networks. Note that we always compare simulations with the same initial seed, but we average across various seeds to evaluate model performance across multiple starting points of the dynamics. We obtain very low values for all metrics that, notably, are very close to the values computed only among simulations on the real network, thus quantifying the intrinsic variability of the dynamics. Overall, the epidemic spreading patterns on the model network closely reproduce those observed on the real WAN.

\subsection{Open data calibration}\label{opendata}

Finally, we show how the model performs when informed with openly available data on air traffic, rather than aggregate OD information from OAG.
We first construct an open dataset by combining two independent data sources:
flight records from OpenSky \cite{essd-13-357-2021}
used to extract network topological information, and passenger traffic statistics from Wikipedia used to infer basin strengths (see Methods).
As open data coverage is incomplete, we can collect information only for a restricted set of 739 basins (about 25\% of the total), which is yet representative (45\%) of the total mobility derived from OAG data.
We then compare model networks calibrated on the collected open data with the empirical OAG network for the restricted basin set.

Table~\ref{tab:opendata} contains an overview of the results.
By definition, the model now reproduces strength values from Wikipedia, which are not perfectly aligned with OAG values. Also, because of the smaller sample size leading to a higher statistical error, structural reconstruction performance slightly deteriorates overall with respect to the OAG-informed model (especially at the level of individual link weights), but recall stays as high as 0.58 and large-scale topological properties of the network remain well reproduced.
In terms of epidemic simulations, the gap between the intrinsic variability for empirical data and the difference between model and empirical scenarios increases. However, model values remain generally compatible with empirical quantities (note that RMSE[$i_m$] and L2[$H$] are sensitive to the cardinality of the node set). Hence, the model remains accurate even for imperfect input information on aggregate OD flows.

\section{Discussion}

In this paper, we introduced a maximum-entropy framework to model origin-destination (OD) flows of the worldwide air mobility network (WAN).
The model extends classical fitness approaches to spatially embedded systems with community structure through a \emph{block-fitness} formulation \cite{bernaschi2022}, in which passenger volumes at individual locations determine connectivity while aggregate passenger fluxes across geographic communities are preserved.
This construction enables the probabilistic generation of synthetic air mobility networks using only coarse-grained information on the system.

The model faithfully reproduces the large-scale topological properties of the empirical WAN, the presence and traffic statistics of individual connections (achieving a link prediction recall of 75\%), as well as the dynamical patterns of epidemic spreading.
The block-fitness approach outperforms alternative formulations that account for the super-linear relation between strengths and degrees \cite{barrat2004architecture} and for the dependence of connection probabilities on geographical distance \cite{bianconi2021information} (see Supplementary Methods for model definitions and Supplementary Note 1 for model comparison).
Our conclusions remain robust when considering global air mobility at a finer geographical resolution, where nodes correspond to individual airports rather than basins of metropolitan areas (details in Supplementary Note 2).

The community partition is specified \emph{a priori} using the UN geoscheme and World Bank regional classification, rather than inferred from network data (see Supplementary Note 3 for another model specification with communities corresponding to ICAO regions).
Hence we do not optimize the model with respect to the chosen partition, and a data-driven refinement of the community structure could further improve performance.
Future work may also develop extensions in which binary and weighted constraints jointly inform the reconstruction \cite{gabrielli2019grand,parisi2020faster}, or evaluate models for their ability to reproduce the features of more realistic contagion models.

Finally, we emphasize that our inference framework is analytic (\emph{white-box}) and therefore complements, rather than competes with, \emph{black-box} machine-learning approaches.
Unlike methods that require access to complete network realizations for training, our framework reconstructs realistic WAN surrogates directly from aggregate information, without ever observing the full network: reconstruction relies only on low-dimensional OD information at the level of individual locations and geographic communities.
Importantly, the model can also be informed with open data on individual flight legs and transit passenger counts, with limited loss of reconstruction accuracy (further information in Supplementary Note 4).
Overall, this work shows that combining degree--strength constraints with coarse-grained spatial organization can be used to generate realistic WAN surrogates for forecasting and policy analysis.

\section{Methods}

\subsection{Model formulation}

Our modeling approach belongs to the class of Exponential Random Graphs (ERGs) \cite{park2004statistical, cover2006elements, bianconi2008entropy,fronczak2018exponential}, which build on the principle of Maximum Entropy \cite{jaynes1957information} to generate ensembles of networks that are maximally random but preserve, on average, some features of the empirical network \cite{squartini2011analytical,squartini2017maximum,cimini2019statistical}.
This approach yields the most unbiased probability distribution describing a network with the imposed features, while model parameters can be efficiently obtained with scalable algorithms \cite{vallarano2021fast}.
For this approach to work in the context of sparse weighted networks, the features to be imposed are both node strengths and degrees \cite{mastrandrea2014enhanced,gabrielli2019grand}.
Indeed, without constraining degrees, the generated network would be fully connected, because strengths alone do not contain information on how many links should be actually realized (see the \emph{irreducibility conjecture} in \cite{cimini2015estimating}): unconstrained entropy maximization distributes the available weight over all possible (in fact all) connections.
When degree information is not available or to reduce computational cost, a heuristic \emph{fitness ansatz} \cite{caldarelli2002scale} can be used to replace the corresponding model parameters with other node properties that are positively correlated with the degree. Typically the strengths themselves are used \cite{garlaschelli2004fitness}, under the assumption that nodes with large strength are likely to have many connections.
The model can then be formulated as a two-step inference procedure \cite{cimini2015systemic,cimini2015estimating}, where links are first sampled independently according to a connection probability that depends on node strengths,
and then assigned a weight according to a \emph{corrected} gravity model prescription.
This recipe thus generates sparse topologies while preserving node strengths.

Here we build on this formulation by further imposing a community structure on the model network while keeping the global strength constraints.
We thus split nodes into a set $\{g\}$ of communities and, for each node $i$, we constrain the strength values computed with respect to each community $g$: $s_{ig} = \sum_{j(\neq i)\in g}w_{ij}$.
The probability of connection for nodes $i$ and $j$ (with $i\neq j$), belonging respectively to community $g_i$ and $g_j$, thus reads:
\begin{equation}
    p_{ij}=\frac{z_{g_ig_j}s_{ig_j}s_{jg_i}}{1+z_{g_ig_j}s_{ig_j}s_{jg_i}}
\label{eq.p_BF}
\end{equation}
which depends on $s_{ig_j}$ and $s_{jg_i}$ (the required inputs of the model).
The free parameter $z_{g_ig_j}$ is determined by setting the expected number of connections among communities $g_i,g_j$ equal to the empirical value:
$\sum_{i\in g_i}\sum_{j\in g_j}p_{ij}=\sum_{i\in g_i}\sum_{j\in g_j}a_{ij}$ (details in Supplementary Methods).
Then, each sampled link $ij$ (realized with probability $p_{ij}$) is assigned the weight:
\begin{equation}
w_{ij} = \frac{s_{ig_j}s_{jg_i}}{p_{ij}W_{g_ig_j}}
\label{eq.w_BF}
\end{equation}
while $w_{ij}=0$ otherwise.
$W_{g_ig_j}=\sum_{i\in g_i}\sum_{j\in g_j}w_{ij}$ is the total weight of connections among nodes in $g_i$ and $g_j$.
The average link weight thus reads $\langle w_{ij}\rangle = s_{ig_j}s_{jg_i}/W_{g_ig_j}$, hence the model preserves, as ensemble averages, the strength by block of each node:
$\langle s_{ig_j}\rangle=\sum_{j\in g_j}\langle w_{ij}\rangle\equiv s_{ig_j}$ (and thus the total strength).
This model formulation can be seen as a composition of fitness models for each block \cite{cimini2015estimating} or as a fitness-based version of discrete latent variable models \cite{karrer2011stochastic,fronczak2013exponential} (see Supplementary Methods for the model formulation tailored for directed networks).

\subsection{Epidemic simulations}

We implement metapopulation simulations of epidemic spreading by coupling a Susceptible-Infected-Recovered (SIR) dynamics taking place in each basin with mobility among different basins according to WAN passenger flows \cite{balcan2009multiscale,BALCAN2010132}.
We use the \emph{Epidemiology 101} package \cite{101}, available at \url{https://github.com/DataForScience/Epidemiology101/tree/master}.
For each basin $m$, the number of susceptible, infected and recovered individuals, denoted $S_m$, $I_m$, and $R_m$ (with $S_m+I_m+R_m=N_m$ the total population $m$), evolves according to a stochastic SIR dynamics, described by the mean-field equations
\begin{equation}
    \begin{cases}
        \dfrac{dS_m}{dt}=-\dfrac{\beta I_mS_m}{N_m}\\[6pt]
        \dfrac{dI_m}{dt}=\dfrac{\beta I_mS_m}{N_m}-\gamma I_m\\[6pt]
        \dfrac{dR_m}{dt}=\gamma I_m
    \end{cases}
\end{equation}
where $\beta$ is the transmission rate and $\gamma$ the recovery rate. The dynamics of the various basins are coupled through the OD travel matrix of the WAN \cite{balcan2009multiscale}, whose link weights determine the passenger flows among basins. Here we set $\pi_{mn}=w_{mn}/(N_m+N_n)$ as the probability that an individual travels from $m$ to $n$ in a given time step.
Simulations start with an initial number of infected $I_0=10$ in a designated seed basin and evolve with a transmission rate $\beta=0.3$ and a recovery rate $\gamma=0.1$, for $300$ time steps.

We select initial seeds as the four most interconnected basins from each continent (Africa, Asia, Europe, North America, South America, and Oceania), resulting in a set of $24$ highly connected basins all over the world: Johannesburg, Cairo, Cape Town, Algiers, Hong Kong, Shanghai, Beijing, Bangkok, London, Paris, Moscow, Milan, New York, Los Angeles, San Francisco, Miami, Sydney, Melbourne, Brisbane, Auckland, Sao Paulo, Rio de Janeiro, Bogota, Santiago.
For each seed, we performed $5$ simulations on the real network and $20$ simulations on different network realizations of the model, for a total of $600$ simulations.

Note that by using OD data to model passenger flows among basins, we are not taking layovers into account within the spreading dynamics.
Indeed, transmission during transit is expected to contribute less to epidemic dynamics than movements between origin and destination populations, for several reasons.
In-flight transmission of respiratory pathogens is relatively uncommon due to high cabin air exchange rates and HEPA filtration \cite{Hertzberg2018}. Moreover, infected travelers are often still in the latent stage during travel, making progression to infectiousness after arrival the dominant mechanism shaping geographical spread. Transit dwell times are also substantially shorter than time spent at the destination.

\subsection{Open data}

\paragraph{Topological information.}
We use the OpenSky COVID-19 Flight Dataset \cite{essd-13-357-2021}, part of the
OpenSky Network's crowdsourced ADS-B sensor data covering global air traffic from January 2019 onward. To match the temporal scope of the OD basin dataset, we restrict our analysis to the period Sept.--Nov. 2019.
Each record in the OpenSky data reports the first and last detected
positions of a given flight; we use these coordinates to approximate the corresponding origin and destination airports, which we then map to their nearest basin. This procedure yields an open-data edge list, from which we derive topological information for each basin.
Note that direct flight connections from OpenSky cannot fully capture the connectivity patterns of OD data, as routes with layovers are decomposed into individual legs. Indeed, degrees inferred from OpenSky are systematically underestimated, therefore we rescale them to match the observed network density of OAG data.

\paragraph{Traffic statistics.}
We employ Wikidata to obtain passenger traffic volumes of all available airports for the period Sept.--Nov. 2019. We map (and sum) these values to the corresponding basins to obtain a proxy of basin strengths.
In order to improve consistency with OD values from OAG, we filter out a handful of small outliers.
Note that since Wikidata counts include transit passengers, obtained strengths are systematically higher than original OD values, but this effect is absorbed by automatic recalibration of model parameters. Lastly, to fully inform the model, we need passenger flows among regions. We obtain them by splitting Wikidata strengths according to rescaled connectivity-by-region obtained from OpenSky.

Supplementary Note 4 contains full details on the procedures described above, as well as results of the different model formulations when calibrated with open data.

\section*{Data availability}
Raw OD data are available under commercial license from OAG.
OpenSky data are available at \url{http://www.opensky-network.org} and Wikidata at \url{https://www.wikidata.org/wiki/Wikidata:Main_Page}.

\section*{Code availability}
Open data and code to implement the network generation models are available at the github repository \url{https://github.com/mnlknt/WAN-fitness-modeling}.

\begin{acknowledgments}
G.~Ca. acknowledges funding from the Italian Ministero dell'Universit\`a e della Ricerca under the programme PRIN 2022 (``re-ranking of the final lists''), number 2022KWTEB7, cup B53C24006470006.
\end{acknowledgments}

\section*{Author contributions}
Designed the study: G.Ci., A.V., and G.Ca.
Curated the OAG data: J.D., A.L.
Collected the Open data: A.M., G.Ci.
Contributed code and repository: G.F., A.M.
Carried out the analysis: G.F., A.M., G.Ci.
Analyzed results: G.F., A.M., G.Ci., J.D., A.L., A.V., G.Ca.
Prepared the manuscript and figures: G.F., A.M., G.Ci.
Revised the manuscript: G.F., A.M., G.Ci., J.D., A.L., A.V., G.Ca.

\section*{Competing interests}
The authors declare no competing interests.

\begin{figure*}[t]
    \centering
    \includegraphics[width=\textwidth]{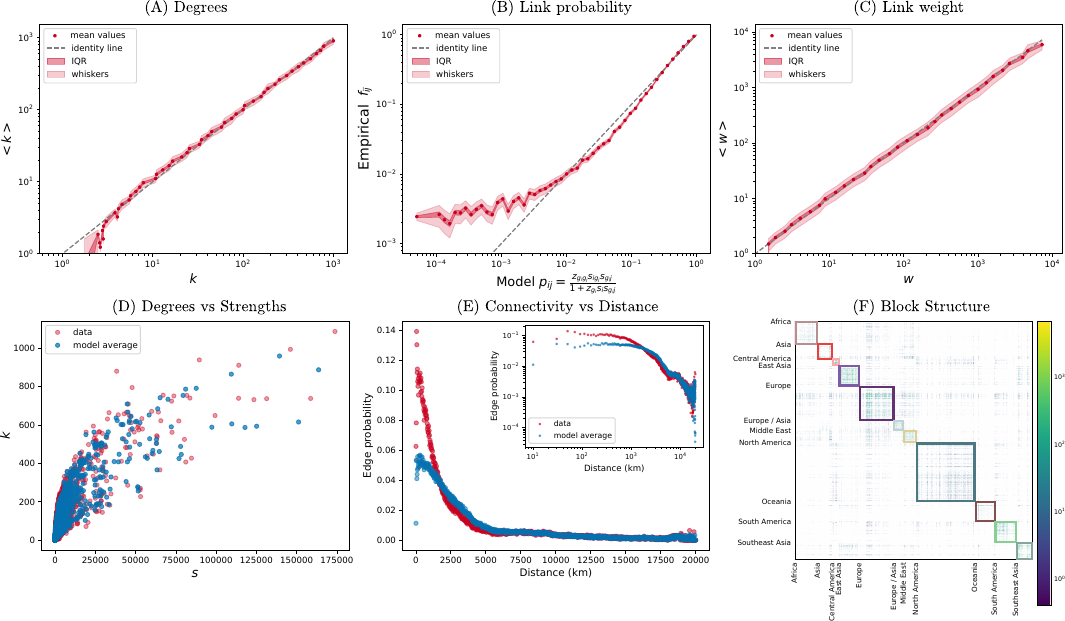}
    \caption{\textbf{Reconstruction of topological properties of the WAN.}
    (A-C) Scatter plots of model vs empirical values (represented by binning abscissa values into 50 uniformly-spaced intervals in log scale). The dashed identity line represents perfect agreement while the shaded regions mark the inter-quartile range [Q1, Q3] and the whiskers of the set of model values (with the lower limit truncated at zero).
    Root Mean Squared Relative Error (RMSRE, definition in Supplementary Table 2) values are computed for binned values to focus on the statistical reconstruction of aggregate trends rather than individual points.
    (A): Model degrees $\avg{k_i} = \sum_{j\neq i} p_{ij}$ versus empirical degrees $k_i$, with RMSRE = 0.18. Quartiles are computed empirically over the ensemble values falling in each bin.
    (B): Empirical link probability (given by the fraction $f_{ij}$ of $ij$ pairs that are actually connected in each bin of $p_{ij}$) as a function of the model link probability $p_{ij}$, with RMSRE = 0.016.
    As each bin contains a single value of $f_{ij}$, quartiles are obtained from the  normal approximation of the binomial distribution mean, with variance $f_{ij}(1-f_{ij})$ divided by the number of $ij$ pairs in the corresponding bin.
    (C): Average model weights $\langle w\rangle_{ij}$ versus empirical weights $w_{ij}$, with RMSRE = 0.063. Quartiles are computed empirically over the ensemble values falling in each bin. For this exercise, we limit the test set to links $w_{ij}>0$ that are actually realized.
    (D-F) Empirical vs model network features.
    (D) Degrees vs strengths relation $s\sim k^\beta$ for individual basins, with OLS fit exponent $\beta=1.2(2)$ on empirical data and $\beta=1.17(0)$ on model data.
    (E) Link probability as a function of distance (binning as in Fig.~\ref{fig:basin_feat}C), with Mean Squared Error (MSE) between empirical and model curve of 1.61.
    (F) Block structure of a sample reconstructed network. Modularity is 0.63 on average over the model ensemble for the imposed partition ($M=0.67$ for real data).}
    \label{fig:rec_basin}
\end{figure*}

\begin{table*}[t]
\caption{Similarity of epidemic simulations on real and model networks, in terms of: area between and L2 distance of prevalence curves; differences of peak time, peak magnitude, epidemic length; RMSE of prevalence vectors; L2 distance of entropy curves (definitions in Supplementary Table 3).
Values on the ``Model'' row are averages over pairwise comparisons of one simulation on the empirical network and one simulation on a model network, while
``Data'' values are computed with simulations on the empirical network only. Uncertainties are given as standard deviations.}
\label{tab:epidemic_model_B_vs_real_variability}
\begin{ruledtabular}
\begin{tabular}{c|ccccccc}
& $\Delta$ area & L2[$i$] & Peak time & Peak magn. & Epid. length & RMSE[$i_m$] & L2[$H$] \\
\hline
Model &
0.56 $\pm$ 0.31 &
0.073 $\pm$ 0.040 &
2.3 $\pm$ 1.8 &
0.0048 $\pm$ 0.0035 &
4.5 $\pm$ 3.2 &
0.050 $\pm$ 0.006 &
14.5 $\pm$ 6.1 \\
Data &
0.43 $\pm$ 0.32 &
0.058 $\pm$ 0.042 &
2.0 $\pm$ 1.7 &
0.0043 $\pm$ 0.0034 &
2.2 $\pm$ 2.2 &
0.024 $\pm$ 0.005 &
13.1 $\pm$ 6.1 \\
\end{tabular}
\end{ruledtabular}
\end{table*}

\begin{table*}[t]
\caption{Model calibrated on open data. Top: Reconstruction of topological properties (same quantities as in Fig.~\ref{fig:rec_basin}).
For empirical filtered data, $\beta=1.2(4)$ and $M=0.26$.
Bottom: Similarity of epidemic simulations on real and model networks (same quantities as in Table~\ref{tab:epidemic_model_B_vs_real_variability}).}
\label{tab:opendata}
\begin{ruledtabular}
\begin{tabular}{c|ccccc|ccc}
         	&	 RMSRE$_s$    &	 RMSRE$_k$    &	 RMSRE$_p$    &	 RMSRE$_w$    &	 Recall    & $\beta$ 	& MSE$_{l(d)}$ 	& $M$ \\
\hline
Model	&	8.36	&	0.28	&	0.057	&	4.53	&	0.58	&	1.3(8)	&	0.56	&	0.18	\\
\end{tabular}
\end{ruledtabular}
\vspace{0.5em}
\begin{ruledtabular}
\begin{tabular}{c|ccccccc}
& $\Delta$ area & L2[$i$] & Peak time & Peak magn. & Epid. length & RMSE[$i_m$] & L2[$H$] \\
\hline
Model &
1.2 $\pm$ 0.5 &
0.17 $\pm$ 0.07 &
4.3 $\pm$ 2.4 &
0.0071 $\pm$ 0.0049 &
7.2 $\pm$ 2.4 &
0.0086 $\pm$ 0.0010 &
13.2 $\pm$ 2.8 \\
Data &
0.55 $\pm$ 0.34 &
0.076 $\pm$ 0.047 &
1.7 $\pm$ 1.4 &
0.0042 $\pm$ 0.0039 &
3.3 $\pm$ 2.7 &
0.0055 $\pm$ 0.0004 &
7.4 $\pm$ 3.6 \\
\end{tabular}
\end{ruledtabular}
\end{table*}

\bibliographystyle{apsrev4-2}
\bibliography{bibliography}

\end{document}